\documentclass[aps,prl,epsfig,twocolumn,floatfix]{revtex4}
\usepackage{epsfig}
\begin{document}

%\tighten \preprint{ \vbox{ \hbox{July 2004} \hbox{...} }}

\title{Local Duality and Charge Symmetry Violation in Quark Distributions}

\author{F. M. Steffens and K. Tsushima}

\affiliation{   NFC - FCBEE - Universidade Presbiteriana Mackenzie,
            Rua da Consola\c{c}\~ao 930, 01302-907,
            S\~ao Paulo, SP, Brazil \\
            IFT - UNESP,
            Rua Pamplona 145, 01405-900,
            S\~ao Paulo, SP, Brazil}

%\maketitle

\begin{abstract}
We use local quark-hadron duality to calculate the nucleon
structure function as seen by neutrino and muon beams. Our result
indicates a possible signal of charge symmetry violation at the
parton level in the very large $x$ region.
\end{abstract}

\maketitle

%\vspace*{1cm}
%%%%%%%%%%%%%%%%%%%%%%%%%%%%%%%%%%%%%%%%%%%%%%%%%%%%%%%%%%%%%%%%%%%%%%%%%%%
There has been a significant activity in the study of local
quark-hadron duality in the last four years, most of it triggered
by high quality data for the proton structure function, at the
resonance region, obtained in the late 1990´s at the JLAB
\cite{jlab2000}. From this experimental result, it was possible to
test quantitatively the Bloom and Gilman ideas \cite{bg70} on the
relation between the exclusive cross section at low $Q^2$ and the
inclusive cross section at high $Q^2$. Specifically, the data from
JLAB shows that the equivalence between $F_2$ calculated from
electron quark scattering, and $F^{res}_2$ calculated from
averaging the resonance structure function, including the nucleon
pole, holds for $Q^2$ as low as $0.5\; GeV^2$.

The QCD justification for local duality was supplied by de
R\`{u}jula, Georgi and Politzer \cite{RGP77}. Using the Nachtmann
variable, $\xi = 2x/(1 + \sqrt{1 + 4 x^2M^2/Q^2})$, they showed
that the lower moments of a structure function $F(\xi, Q^2)$ are
independent of $Q^2$, up to perturbative QCD corrections, in the
resonance and in the scaling region. For higher moments, however,
higher twist contributions are fundamental. As the lower moments
give the most important contributions when reconstructing the
$\xi$ (or $x$) dependence of the structure functions, it follows
that $F^{res} (x, Q^2) \approx F (\xi)$. In reality what happens
is that $F^{res} (x, Q^2)$ oscillates around the scaling function
as we approach the resonance poles, the origin of these
oscillations being higher twists contributing with alternated
signs. However, these higher twist contributions cancel on the
average. The scaling function, on the other hand, is calculated at
very high $Q^2$, meaning that all the resonance peaks have moved
to the large $x$ region. Following \cite{bg70,RGP77,wally2001} we
relate the scaling structure function to the elastic part of the
structure function calculated at the nucleon pole. This enables us
to estimate the behavior of the scaling structure function in the
large $x$ region, as long as local duality holds.

Our particular interest is the iso-scalar $F^{\nu N}_2$ structure
function measured in deep inelastic neutrino-nucleon scattering.
At large $x$, assuming that charge symmetry holds at the parton
level, we should have in leading order:

\begin{equation}
F_2^{\nu N} (x \rightarrow 1) \simeq x[u(x) + \overline{u}(x) +
d(x) + \overline{d}(x)], \label{e1}
\end{equation}
where we did not write the contribution from the strange quarks
because they are not expected to contribute in this region
\cite{par,bar}. On the other hand, the iso-scalar structure
function $F^{\mu N}_2$, measured in muon scattering should be
given, at large $x$, by:

\begin{equation}
F_2^{\mu N} (x \rightarrow 1) \simeq \frac{5}{18} x [u(x) +
\overline{u}(x) + d(x) + \overline{d}(x)]. \label{e2}
\end{equation}
In fact, both neutrino and muon structure functions of an
iso-scalar target have been measured already, but at intermediate
$x$ \cite{data,yang2001}. In this region the target mass
corrections, along with the nucleon strange and anti-strange quark
distributions, are essential to reconcile both experiments with
the assumption of universal parton distributions
\cite{yang2001,boros}. With the absence of strange
quarks and antiquarks, we must have from Eqs. (\ref{e1}) and
(\ref{e2}) that:

\begin{equation}
\frac{5}{18} F_2^{\nu N}(x \rightarrow 1) \simeq F_2^{\mu N} (x
\rightarrow 1) \label{e3}
\end{equation}

A failure of equation (\ref{e3}) would suggest that either an
unexpected strange distributions, perhaps intrinsic strangeness
\cite{brod80}, at large $x$ or that charge symmetry between the
proton and the neutron affects the large $x$ distributions
\cite{rodionov}. Any of these two conclusions are very significant
and justify a deeper study of the structure functions as probed by
neutrinos and muons in this region. This is the main objective of
this letter. We will use local quark-hadron duality to investigate
relation (\ref{e3}).

For this purpose, we will need the hadronic tensor that enters
in the quasi elastic neutrino-nucleon cross section,
$\nu_\mu(\overline{\nu}_\mu) + n (p) \rightarrow \mu^- (\mu^+) + p
(n)$. Keeping only the relevant and sufficient terms for this
study \cite{tonybook}, we start with the matrix elements of the
charged current, which is given by:

\begin{eqnarray}
<p(P^\prime)|J_+^\mu (0) |n(P)> &=& <n(P^\prime)|J_-^\mu (0)
|p(P)> \nonumber \\*
&&\hspace{-3cm}=\overline{u}(P^\prime)\left[F_1^V(Q^2)\gamma^\mu +
\frac{i\sigma^{\mu\nu}q_\nu}{2M} F_2^V(Q^2) \right. \nonumber \\*
& & \left. - G_A(Q^2)\gamma^\mu\gamma_5\right]u(P), \label{e4}
\end{eqnarray}
where $F_1^V (Q^2)$ and $F_2^V (Q^2)$ are, respectively, the
iso-vector Dirac and Pauli form factors, and $G_A (Q^2)$ the axial
form factor. The elastic part of the hadronic tensor calculated
from charged current is then:

\begin{equation}
W_{\mu\nu}^{el} = -F^{el}_1 \frac{g_{\mu\nu}}{M} + F^{el}_2
\frac{P_\mu P_\nu}{2 M^3 \tau} + i F^{el}_3
\varepsilon_{\mu\nu\alpha\beta}\frac{P^\alpha q^\beta}{4 M^3
\tau}, \label{e5}
\end{equation}
with $\tau = Q^2/4M^2$ and

\begin{equation}
F^{el}_1 = \frac{M}{2}\delta\left(\nu -
\frac{Q^2}{2M}\right)[\tau(G^V_M)^2 + (1 + \tau)G_A^2], \label{e6}
\end{equation}

\begin{equation}
F^{el}_2 = M \tau \delta\left(\nu -
\frac{Q^2}{2M}\right)\left[\frac{(G^V_E)^2 + \tau (G^V_M)^2}{1 +
\tau} + G_A^2\right], \label{e7}
\end{equation}

\begin{equation}
F^{el}_3 = M \tau \delta\left(\nu - \frac{Q^2}{2M}\right)[2 G^V_M
G_A], \label{e8}
\end{equation}
where $\nu$ in Eqs. (\ref{e6})-(\ref{e8}) is the energy transfer
between the beam and the target. The iso-vector electric and
magnetic form factors are given by $G^V_{E,M} = G^p_{E,M} -
G^n_{E,M}$, with $G^N_E = F^N_1 - \tau F^N_2$ and $G^N_M = F^N_1 +
F^N_2$.

As previously discussed, local quark-hadron duality is translated
into approximately equal low moments for the resonance and the
scaling structure functions for each resonance. Hence, looking at
the nucleon pole only, we will have the following equation
relating the scaling structure function $F^{\nu N}_2$ and the
elastic contribution $F_2^{el}$:

\begin{equation}
\int_{\xi_{th}}^1 F_2^{\nu N} (\xi) d\xi \simeq \int_{\xi_{th}}^1
F^{el}_2 (\xi, Q^2) d\xi, \label{e9}
\end{equation}
where $\xi_{th} = 2 x_{th}/(1 + \sqrt{1 + x_{th}^2/\tau})$ is the
Nachtmann variable at the pion threshold, with $x_{th} = Q^2/(Q^2
+ m_\pi (2M + m_\pi))$. The $Q^2$ independence of the scaling
$F_2^{\nu N}$ means that we are not taking into account the
perturbative QCD corrections to it. We now use Eq. (\ref{e7}) on
the right hand side of Eq. (\ref{e9}):

\begin{equation}
\int_{\xi_{th}}^1 F_2^{\nu N}(\xi) d\xi \simeq \frac{\xi_0^2}{4 -
2\xi_0} \left[\frac{(G^V_E)^2 + \tau (G^V_M)^2}{1 + \tau} +
G_A^2\right], \label{e10}
\end{equation}
where $\xi_0$ is the Nachtmann variable at the nucleon pole.
Taking the derivative of Eq. (\ref{e10}) with respect to $x_{th}$,
with $\xi_0$ fixed, we get:

\begin{eqnarray}
F_2^{\nu N}(x_{th}) &\simeq& -\beta \left[\frac{(G_M^V)^2 -
(G_E^V)^2}{4M^2 (1 + \tau)^2} \right. \nonumber \\* &+&
\left.\frac{1}{1 + \tau} \left( \frac{d(G_E^V)^2}{dQ^2} + \tau
\frac{d(G_M^V)^2}{dQ^2}\right) + \frac{dG_A^2}{dQ^2}\right], \nonumber \\*
\label{e11}
\end{eqnarray}
with $\beta = (Q^4/M^2)(\xi_0^2/\xi_{th}^3)(2 -
\xi_{th}/x_{th})/(4 - 2\xi_0)$. The same calculation for the
$F_1^{\nu N}$ and $F_3^{\nu N}$ structure functions gives:

\begin{eqnarray}
F_1^{\nu N}(x_{th}) &\simeq& -\frac{\beta}{2}
\left[\frac{-G_A^2}{4M^2\tau^2} + \frac{d(G_M^V)^2}{dQ^2} + \frac{1
+ \tau}{\tau} \frac{dG_A^2}{dQ^2}\right], \nonumber \\* F_3^{\nu
N}(x_{th})&\simeq& -\beta\frac{d(2G_M^V G_A)}{dQ^2}. \label{e12}
\end{eqnarray}
Notice that $F_1^{\nu N}(x_{th})$ and $F_2^{\nu N}(x_{th})$ have
the same behavior in the $\tau\rightarrow\infty$ region. $F_3^{\nu
N}(x_{th})$, on the other hand, is associated with an interference
between the vector and axial parts of the charged current.
Finally, a similar calculation can be made for the eletromagnetic
structure functions. We quote here our result for the $F_2^{\mu
p}$ case:

\begin{eqnarray}
F_2^{\mu p}(x_{th}) &\simeq& - 2\beta\left[\frac{(G_M^p)^2 -
(G_E^p)^2}{4M^2(1 + \tau)^2} \right. \nonumber \\* &+& \left.
\frac{1}{1 + \tau} \left(\frac{d(G_E^p)^2}{dQ^2} + \tau
\frac{d(G_M^p)^2}{dQ^2}\right)\right], \label{e13}
\end{eqnarray}
which agrees with \cite{wally2001}, except for a $1/x_{th}$ in the
$\beta$ factor and an overall minus sign. As only ratios are shown
in \cite{wally2001}, the conclusions presented in that work are
unaffected.

Using Eqs. (\ref{e11}) and (\ref{e13}) in Eq. (\ref{e3}) we have,
in the large $Q^2$ limit, that:

\begin{eqnarray}
& &\frac{5}{18} F_2^{\nu N}(x=x_{th} \rightarrow 1) - F_2^{\mu
N}(x=x_{th} \rightarrow 1) \simeq \nonumber \\* &+&
\frac{13}{18}\beta\left(\frac{d(G_M^p)^2}{dQ^2} +
\frac{d(G_M^n)^2}{dQ^2}\right) + \frac{5}{9}\beta\frac{d(G_M^p
G_M^n)}{dQ^2} \nonumber \\* &-&
\frac{5}{18}\beta\frac{dG_A^2}{dQ^2}, \label{e14}
\end{eqnarray}
which is not zero, unless some numerical coincidence happens. Eq.
(\ref{e14}) incorporates the main point of this work. To
understand how large the deviation of $5F_2^{\nu N}/18F_2^{\mu
N}$ is from 1, when using quark-hadron duality to calculate the
scaling functions, we used a world data parametrization
\cite{formfactor} to calculate the form factors appearing in
(\ref{e11}) and (\ref{e13}). The result, shown in Figure 1, is
clearly different from 1. 
The calculation should not be trusted
for $x = x_{th} \lesssim 0.78 $, where $Q^2 \lesssim 1\;GeV^2$.
However, at $x = x_{th} \sim 0.9$, $Q^2 \sim 2.5 \; GeV^2$ and
$W^2 \sim 1.25\; GeV^2$, a region where local quark - hadron
duality has more chances of being respected, although its validity,
mainly for $Q^2 < 1.5 \; GeV^2$, is still controversial \cite{ricco,dong}.
In any case, according to Ref. \cite{wallynpa} the extraction of the
elastic form factors from the scaling structure functions gives a reasonable 
agreement with the experimental data, indicating that the calculation
of the large $x$ structure functions, in the present kinematical regime,
may be justified at some extent. Finally, another source of error that
could affect the result of Fig. 1 is the experimental uncertainty
on the elastic form factors, which are found to be around $10 \; \%$ in the
$Q^2 \sim 2 \; GeV^2$ region \cite{budd}.

\begin{figure}[htb]
\epsfig{file=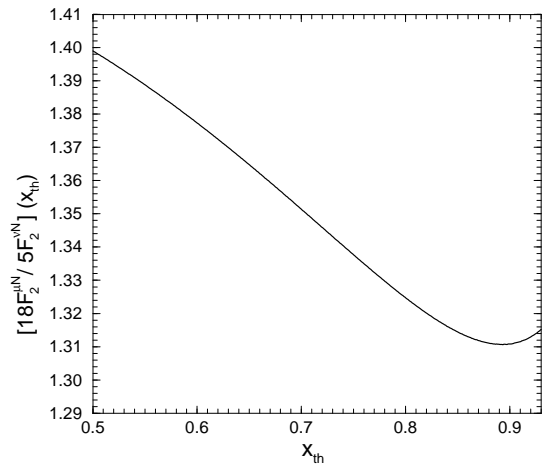,height=7cm,angle=-90} \caption{The
ratio between the iso-scalar structure function as probed by
neutrino and muon beams. In leading order QCD, this ratio should
approach 1 as $x=x_{th}\rightarrow 1$} \label{fig1}
\end{figure}

The effect shown in Figure 1 is larger than the known
limitations of local quark - hadron duality. If taken seriously,
they show an effect at large $x$ that is not marginal. To
understand it, let's look at the following ratio between the
scaling functions, where it is assumed that there is a charge
symmetry violation at the parton level \cite{boros1998}:

\begin{eqnarray}
R_c(x) &\equiv& \frac{F_2^{\mu N}(x)}{\frac{5}{18}F_2^{\nu N}(x) -
x[s(x) + \overline s(x)]/6} \nonumber \\* 
&\approx& 1 - \frac{s(x) - \overline s(x) (x)}{\overline Q(x)} \nonumber \\* 
& & + \frac{4\delta u(x) - \delta\overline u(x) - 4 \delta d(x) +
\delta\overline d(x)}{5\overline Q(x)},
\label{e15}
\end{eqnarray}
where $\overline Q(x) = \sum_{q=u,d,s}[q(x) + \overline q(x)] - 3
[s(x) + \overline s(x)]/5$, and the charge symmetry breaking terms
are $\delta u(x) = u^p(x) - d^n(x)$, $\delta d(x) = d^p(x) -
u^n(x)$, and similar for the anti-quarks. In the large $x$ region,
any charge symmetry breaking coming from anti-quarks should be
negligible, and assuming that the strange and anti-strange
distributions do not contribute, we will have:

\begin{eqnarray}
R_c(x\rightarrow 1) &\simeq& \frac{F_2^{\mu N}(x\rightarrow
1)}{\frac{5}{18}F_2^{\nu N}(x\rightarrow 1)} \nonumber \\*
&\approx& 1 + \frac{4(\delta u(x\rightarrow 1) -  \delta
d(x\rightarrow 1))}{5\overline Q(x\rightarrow 1)}.
\label{e16}
\end{eqnarray}
Therefore, we can explain the results encapsulated in Eq.
(\ref{e14}) and in Figure 1, if we use Eq. (\ref{e16}) and 
allow for charge symmetry
breaking in the quark distributions in the large $x$ region. Our
result requires that $\delta u(x\rightarrow 1) > \delta
d(x\rightarrow 1)$, which is the same sign as obtained in the bag
model calculation of Rodionov et al. \cite{rodionov} in the very
large $x$ region, although vanishingly small, as this theoretical
calculation predicts a significant effect at intermediate $x$
only.

It is also useful to look at the size of the effect given by Eq.
(\ref{e14}) relative to the total magnitude of the structure
functions, $2[(5/6) F_2^{\nu N}(x\rightarrow 1) - 3 F_2^{\mu N}
(x\rightarrow 1)]/((5/6) F_2^{\nu N}(x\rightarrow 1) + 3 F_2^{\mu
N} (x\rightarrow 1))$. This is shown in Figure 2.
\begin{figure}[htb]
\epsfig{file=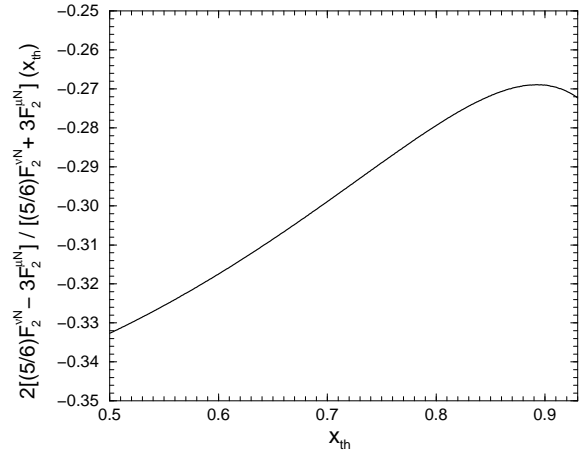,height=7cm,angle=-90} \caption{The
difference between $F_2^{\nu N}(x_{th})$ and $F_2^{\mu N}(x_{th})$
normalized by their total contribution.} \label{fig2}
\end{figure}
As before, there is a definite signal in the region around $x =
x_{th} \sim 0.9$, of about 27 $\%$, which indicates a possible
charge symmetry breaking in the quark distributions even if we
allow a 20 $\%$ error coming from the uncertainty in the local quark-hadron
duality relations and the experiemental determination of the 
elastic form factors. Of course, our assumption of a vanishing
strange distributions at around $x \sim 0.9$ may be questionable.
However, fits of the world data, including the ones that allow for
an asymmetric strange distribution at large $x$, corroborate this
assumption \cite{bar,nutev}.

The charge symmetry violation in the large $x$ valence quark
distributions calculated here can have significant effects in
other areas of particle physics. For instance, recently Londergan
and Thomas \cite{lon2004} analyzed the impact of such 
violation on the determination of the Weinberg angle as
measured by NuTeV \cite{nutev2}. According to their analysis and
our Figure 1, the NuTeV anomaly would become larger because we
have $R_c (x\rightarrow 1) - 1 > 0$, although nuclear corrections
still have to be taken into account. In any case, the need for
physics beyond the standard model would be more pressing. In
summary, we have used local quark-hadron duality to study the
relation between the iso-scalar structure function as probed by
neutrino and muon beams. Our result indicates a possibly sizeable 
violation of charge symmetry in the valence quark distributions at very
large $x$.

%%%%%%%%%%%%%%%%%%%%%%%%%%%%%%%%%%%%%%%%%%%%%%%%%%%%%%%%%%%%%%%%%%%%%%%%%%
\acknowledgments

We would like to thank Wally Melnitchouk for discussions on local
quark-hadron duality, and to Tony Thomas for a reading of the
manuscript. This work was supported by FAPESP (03/06814-8,03/10754-0),
CNPq (308932/2003-0) and MackPesquisa .

%%%%%%%%%%%%%%%%%%%%%%%%%%%%%%%%%%%%%%%%%%%%%%%%%%%%%%%%%%%%%%%%%%%%%%%%%%

%%%%%%%%%%%%%%%%%%%%%%%%%%%%%%%%%%%%%%%%%%%%%%%%%%%%%%%%%%%%%%%%%%%%%%%%%%

\end{document}